               \documentclass[preprint,review,12pt, authoryear]{elsarticle}

 \usepackage{graphics}

\usepackage[usenames,dvipsnames,svgnames,table]{xcolor} 
		\usepackage{tabu}
		\usepackage{longtable}
		\usepackage{afterpage}
		\usepackage{longtable,ltxtable,booktabs}
		\usepackage{amssymb}

		\usepackage[nodots,nocompress]{numcompress}

		\usepackage{lineno}
		\usepackage[letterpaper,pdftex,colorlinks,bookmarks,citecolor=black,%
  bookmarksnumbered,bookmarksopen=true,breaklinks,%
  hyperfootnotes=false,linkcolor=black]{hyperref}
		\newif\ifincludegraphics
		\includegraphicstrue 
		\usepackage{epstopdf}
		\usepackage{multirow}
		\usepackage{tabularx}
		\usepackage{amsmath}

\biboptions{super}

\journal{Icarus}

\newcommand\micron{$\mu$m}
\newcommand\eg{\textit{e.g.}}
\newcommand\methane{CH$_{4}$}
\newcommand\nitrogen{N$_{2}$}
\newcommand\phaseNCH{$\bf{\overline{N_{2}}}$:CH$_{4}$}
\newcommand\phaseCHN{$\bf{\overline{CH_{4}}}$:N$_{2}$}
\newcommand\NCH{N$_{2}$:CH$_{4}$}
\newcommand\CHN{CH$_{4}$:N$_{2}$}
\begin{document}

\begin{frontmatter}



\title{Absorption Coefficients of the Methane-Nitrogen Binary Ice System:
Implications for Pluto}


\author[label1]{S. Protopapa}
\author[label2]{W.M.~Grundy}
\author[label3]{S.C.~Tegler}
\author[label4]{J.M.~Bergonio}

\address[label1]{University of Maryland, Department of Astronomy, College Park, MD, United States}
\address[label2]{Lowell Observatory, Flagstaff, AZ, United States}
\address[label3]{Department of Physics and Astronomy, Northern Arizona University, Flagstaff, AZ, United States}
\address[label4]{Department of Physics and Astronomy, University of Hawai'i, Manoa Honolulu, HI, United States}

\begin{abstract}
The methane-nitrogen~phase diagram of \citet{Prokhvatilov1983} indicates that at temperatures relevant to the surfaces of icy dwarf planets like Pluto, two phases contribute to the methane absorptions: nitrogen saturated with methane \phaseNCH~and methane saturated with nitrogen \phaseCHN. No optical constants are available so far for the latter component limiting construction of a proper model, in compliance with thermodynamic equilibrium considerations.
New optical constants for solid solutions of methane diluted in nitrogen~(\NCH) and nitrogen diluted in methane (\CHN) are presented
at temperatures between 40 and 90~K, in the wavelength
range 1.1--2.7 $\mu$m~at different mixing ratios. These
optical constants are derived from transmission measurements of
crystals grown from the liquid phase in closed cells.  A systematic study of the changes of  methane and nitrogen solid mixtures spectral behavior with mixing ratio and temperature is presented.
\end{abstract}


\end{frontmatter}

\section{Introduction}
Pluto, Eris, and Makemake, unlike most trans-Neptunian objects (TNOs)
with water-ice rich or featureless surfaces \citep{Barucci2008}, display
infrared spectra dominated by methane ice \citep{Brown2008}.  These three
TNOs are often compared with Neptune's large satellite
Triton, since its spectrum is dominated by methane ice and it
is thought to have formed similarly to Pluto, Eris, and Makemake, prior
to its capture into a retrograde orbit around Neptune.  In addition to
methane ice, nitrogen ice has been directly detected on Pluto and Triton
via the 2.148-$\mu$m absorption band \citep{Cruikshank1984,Owen1993}.
Additional evidence for nitrogen ice on Pluto and Triton comes from shifts
of their methane absorption bands to shorter wavelengths, which occurs, as described
by \citet{Quirico1997}, when methane is dissolved at low concentrations in
a matrix of solid nitrogen.  Moderately high signal-to-noise spectra of
Makemake and Eris show no evidence for the presence of the 2.148-$\mu$m
nitrogen absorption feature.  Their methane bands do present subtle
shifts, albeit smaller than the shifts measured in spectra of Pluto
and Triton.  A laboratory study by \citet{Brunetto2008} showed that
smaller shifts correspond to higher methane abundances.  The lack of the
2.15-$\mu$m nitrogen absorption band and the smaller methane wavelength
shifts led several authors \citep{Brown2007,Alvarez-Candal2011,Merlin2009}
to the conclusion that Eris and Makemake are not nitrogen dominated,
contrary to Pluto and Triton.

Thermodynamic equilibrium dictates that if methane and nitrogen ices
are both present, for most of the range of possible nitrogen/methane
relative abundances, two distinct phases must coexist at
temperatures relevant to the surfaces of these icy dwarf planets
\citep{Prokhvatilov1983,Lunine1985}: methane ice saturated with nitrogen
and nitrogen ice saturated with methane.

\citet{Tegler2010} demonstrated that the depth of the methane and
nitrogen absorption bands and the wavelength shift of the observed
methane absorption features should not be used as proxy for the
methane-nitrogen mixing ratio.  The phase composition as dictated by
thermodynamic equilibrium must be taken into account.  In particular,
\citet{Tegler2010} model each of the observed methane absorption bands
with a binary mixture of methane ice saturated with nitrogen and nitrogen
ice saturated with methane, according to the methane-nitrogen phase
diagram of \citet{Prokhvatilov1983}. This technique, applied so far
to the cases of Eris and Pluto only, results in the finding that the
bulk volatile composition of Eris is similar to that of Pluto, with both
objects being dominated by nitrogen ice. A correct modeling of the methane
absorption bands has strong implications not only on the methane-nitrogen
mixing ratio,  but also enables exploration
of stratification properties
as well as heterogeneity of these targets \citep{Tegler2010,Tegler2012}. Because of the lack of absorption coefficients for methane saturated
in nitrogen and for nitrogen saturated in methane at the appropriate
temperatures in the visible wavelength ranges, \citet{Tegler2010} approximate the methane-dominated phase
by using pure methane absorption coefficients and the highly diluted phase
by shifting pure methane coefficients by amounts seen for highly diluted
samples \citep{Quirico1997}. 

In this paper we provide optical constants in the wavelength range 1.1--2.7~$\mu$m of solid solutions of methane
diluted in nitrogen, \NCH, and nitrogen diluted in methane, \CHN, at temperatures
between 40 and 90~K and at different mixing ratios (\url{http://www2.lowell.edu/users/grundy/abstracts/2015.CH4+N2.html}), allowing a proper
model to be constructed for any TNO where the methane/nitrogen ratio
falls between the two solubility limits such that both saturated phases are present. Gaining more detailed knowledge of the methane/nitrogen mixing ratio and
phase state of Pluto, Triton, Eris, and Makemake will enable a better understanding of the processes responsible for volatile loss
and retention on TNOs \citep{Schaller2007}.  It will also help constrain
the seasonal behaviors of their atmospheres, supported by vapor pressure
equilibrium with surface ices.  It will also shed light on the photolytic
and radiolytic chemistry that can occur within the surface ices, since
radicals produced by energetic radiation will encounter and react with
different molecules within a nitrogen-dominated or a methane-dominated
solid phase.
\section{Laboratory Experiments}
The experiments reported here were conducted in a new laboratory ice facility located in the Department of Physics and Astronomy of Northern Arizona University. A detailed description of this facility is given by \citet{Tegler2010,Tegler2012} and \citet{Grundy2011}. We used the closed cell technique for the ice sample preparation.  This technique consists in the growth of crystals from the liquid phase in a closed cryogenic cell. Crystals are grown as follows. The sample is prepared in gas form in a 2 liter mixing volume. Here we report experiments obtained mixing methane (\methane) and nitrogen (\nitrogen). The purities of the gases used are 99.999\% for
\methane~and 99.9\% for \nitrogen, as reported by the vendors. The gas is set to flow into an empty cell, which is at a temperature slightly higher than the melting point of the ice sample. In the case of mixtures, the melting point depends on the mixing ratio of the gases and we use, as reference, the \methane-\nitrogen~phase
diagram of \citet{Prokhvatilov1983}. The gas, once in the cell, condenses immediately to liquid. The liquid is frozen by reducing the temperature in the cell at a rate of 0.1~K~minute$^{-1}$. A thermal gradient is maintained within the cell with the top and bottom heaters such that the crystal grows from the bottom to the top. Once the sample is frozen, the vertical thermal gradient is minimized. After the initial ice spectrum is recorded, the temperature is ramped down
at 0.1~K~minute$^{-1}$. We recorded spectra at temperatures between 40 and 90~K.

\begin{figure}
	\ifincludegraphics
 	\centering
	\includegraphics[width=0.41\textwidth]{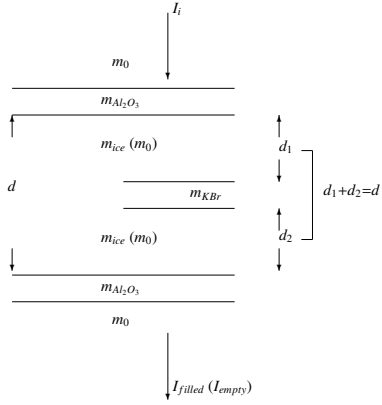}
	\fi
	\caption{Schematic view of the cell interior. See text for details. The incident radiation $I_{i}$ is transmitted through media of different refractive index $m$. $I_{empty}$ and $I_{filled}$ represent the transmitted radiation when the cell is empty (the vacuum $m_{0}$ is between the sapphire windows $m_{Al_{2}O_{3}}$, and the potassium bromide spacer $m_{KBr}$) and filled by the sample ($m_{ice}$), respectively.}
	\label{cell}
\end{figure}

A 5~mm cell, with sapphire (Al$_{2}$O$_{3}$) windows, was used (Figure~\ref{cell}). Thinner samples, needed in the case of mixtures with high \methane~content, were prepared by use of a transparent potassium bromide (KBr) or calcium fluoride (CaF$_{2}$) spacer between the windows. These KBr and CaF$_{2}$ spacers occupied only part of the cell, enabling the choice of the full or reduced thickness, simply by moving the sample relative to the spectrometer beam. The thickness of the sample in absence of spacer is equal to the cell depth, $d=d^{\prime}$, which is known. On the other side of the cell, due to the uncertainties on the depth of the KBr and CaF$_{2}$ spacers, the sample has an unknown thickness, $d=d^{\prime\prime}$. This is computed with a liquid CH$_{4}$ experiment. Two transmission spectra are recorded, $T^{\prime}$ and $T^{\prime\prime}$, by positioning the spectrometer beam away and in correspondence of the spacer, respectively.  From the two transmission spectra, we compute the imaginary part of the refractive index, $k^{\prime}$ and $k^{\prime\prime}$, which will be function of $d^{\prime}$ and the unknown $d^{\prime\prime}$, respectively. Because the imaginary part of the refractive index, $k$, is a property intrinsic to the material, we compute $d^{\prime\prime}$ by setting $k^{\prime}$=$k^{\prime\prime}$ (see Section~\ref{DataAnalysis}).

Spectra were recorded with a Nicolet Nexus 670 Fourier
transform infrared (FTIR) spectrometer, covering the range 2799.9 -- 12000.1~cm$^{-1}$ at a sampling interval
of 0.24~cm$^{-1}$, resulting in a spectral resolution of 0.6~cm$^{-1}$
(FWHM of unresolved lines). We averaged over 100 spectral
scans to improve the signal-to-noise ratio. 
\begin{figure}[!h]
	\ifincludegraphics
 	\centering
	\includegraphics[width=0.48\textwidth]{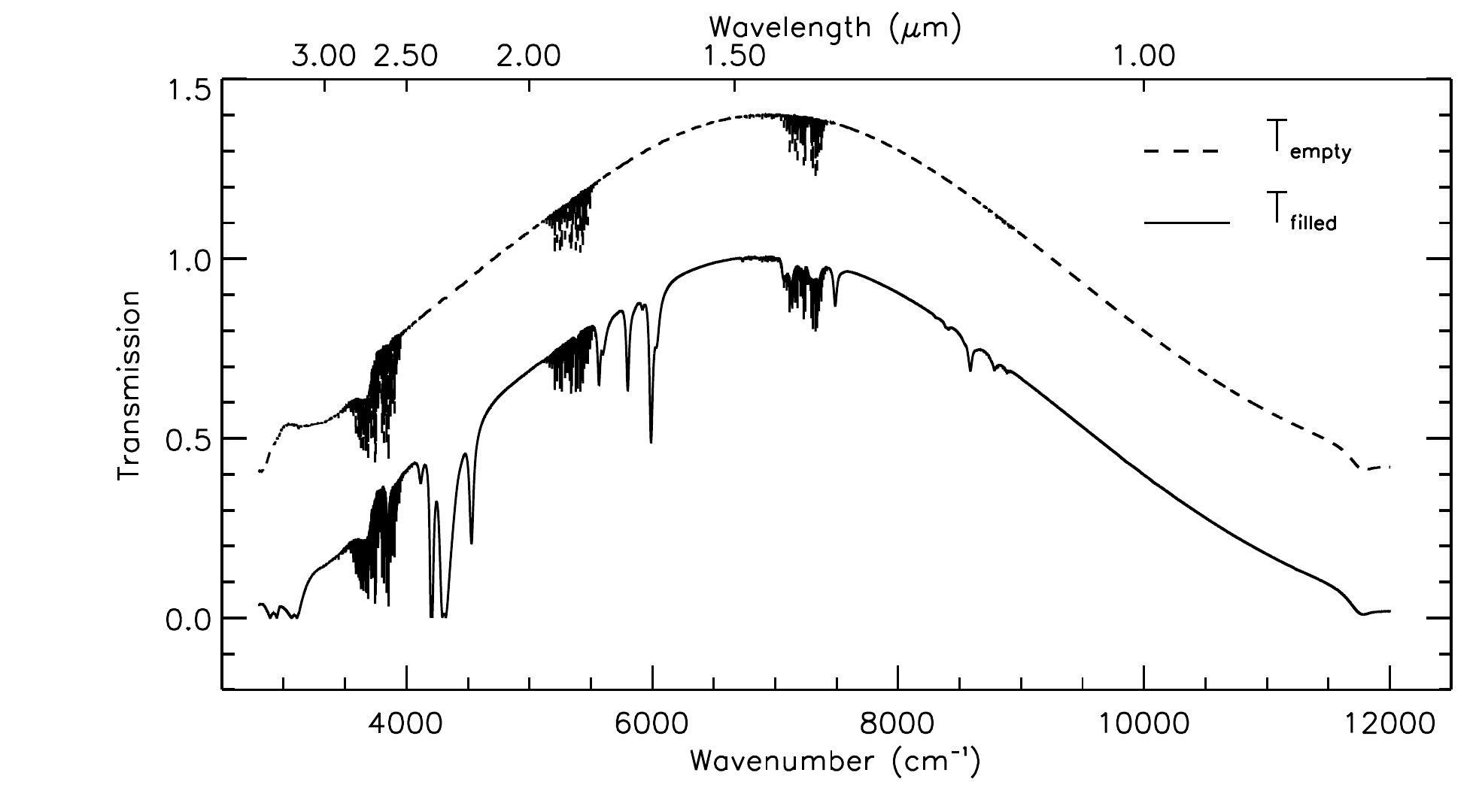}
	\fi
	\caption{The transmission spectra as a function of wavelength and wavenumber, recorded with the cell empty (dashed line) and filled (continuum line) by pure CH$_{4}$ ice at 40K. The upper spectrum has been shifted along the y-axis by 0.4 for clarity.}
	\label{Transmission spectra}%
\end{figure}
\section{Data Analysis}\label{DataAnalysis}
\subsection{Calculation of n and k}
The purpose of our measurements is to compute optical constants of methane-nitrogen ice mixtures, \methane-\nitrogen. The optical constants of a material are the real and imaginary part of the complex refractive index $m(\nu) = n(\nu)+ik(\nu)$, and are both functions of the frequency $\nu$. Let $T(\nu)$ be the transmission spectrum of an ice sample within a cell of thickness $d$ (Figure~\ref{cell}), we compute the imaginary part of the refractive index, $k$, via the Beer-Lambert absorption law:
\begin{equation}
k(\nu) = -\frac{1}{4\pi\nu d}\ln T(\nu).
\end{equation}
In order to remove the flux distribution of the illumination source, the transmission function of the spectrometer and the cell, the detector spectral sensitivity function, and the water vapor contamination \citep{Grundy1998JGR,Tegler2012}, we compute the transmission spectrum $T(\nu)$ as: 
\begin{equation}\label{transmission}
T(\nu)=\frac{T_{filled}(\nu)}{T_{empty}(\nu)},
\end{equation}
where $T_{filled}(\nu)$ and $T_{empty}(\nu)$ represent the transmission through the cell when filled by the sample and empty, respectively (Figure~\ref{Transmission spectra}). Specifically,  $T_{empty}(\nu)$ is the average of the transmission spectra through the empty cold cell recorded before and after the ice sample experiment.
\begin{figure*}
	\ifincludegraphics
 	\centering
	\includegraphics[width=\textwidth]{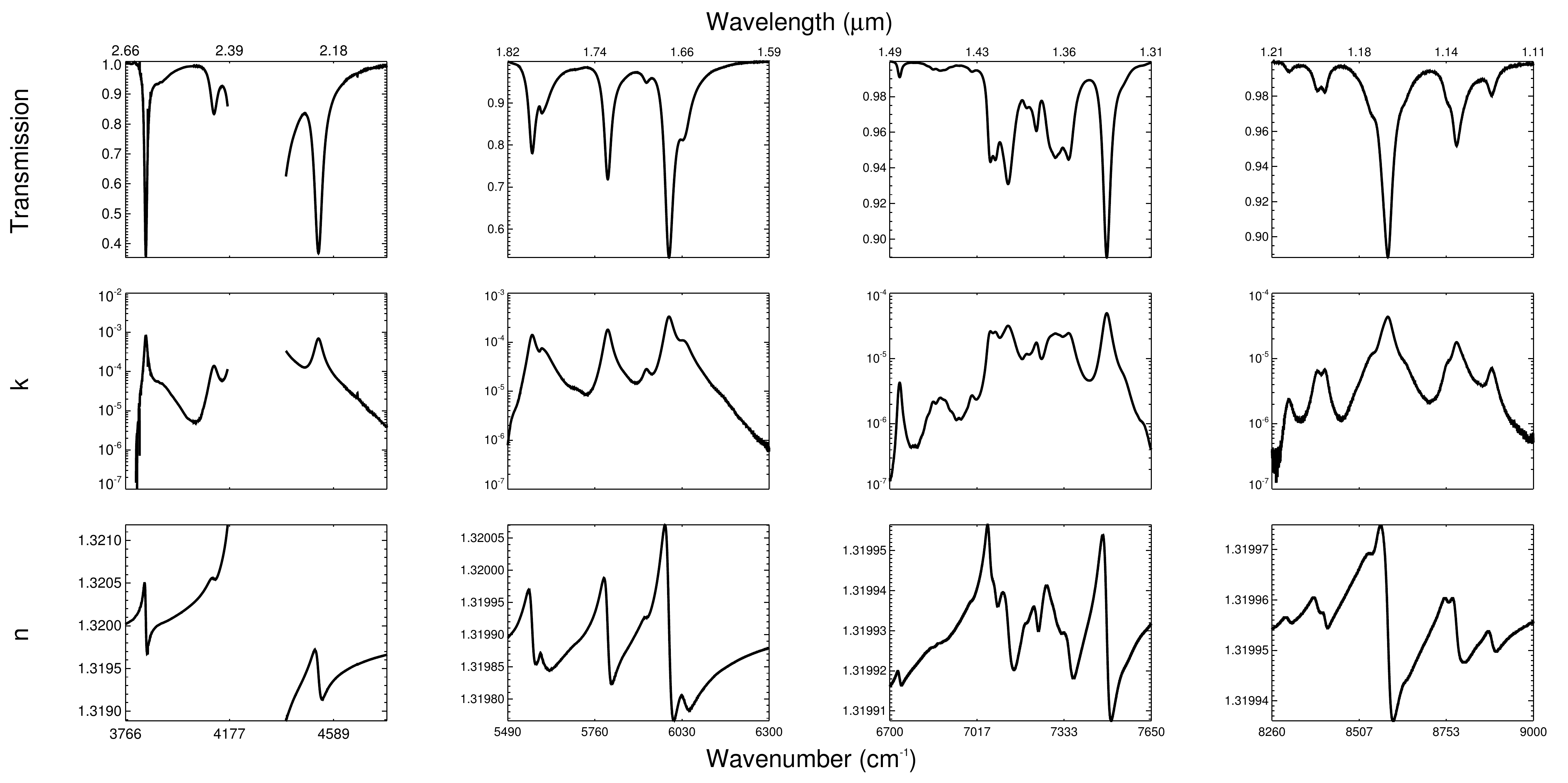}
	\fi
	\caption{Top, middle, and bottom rows show the transmission spectrum, the imaginary part of the refractive index, and the real part of the refractive index of pure CH$_{4}$ at 40K, respectively, in four different bands.}
	\label{OpticalConstants}
\end{figure*}
The transmission spectra, whether the cell is filled or empty, present a high frequency interference pattern due to the windows in the cell, which we remove using a Fourier filter. The resultant transmission spectrum $T(\nu)$ is affected by subtle slopes from a variety of sources \citep{Grundy1998JGR}. In order to remove these slopes, we divide the spectrum by the fit to the continuum regions adjacent to the absorption bands to be analyzed. The continuum of our spectra is well fit by a third order polynomial.

Given the imaginary part of the refractive index, $k(\nu)$, we calculate $n(\nu)$ from the Kramers-Kronig dispersion relation
\begin{equation}
n(\nu)= n_{0}+\frac{2}{\pi}\int\frac{\nu^{\prime} k(\nu^{\prime})}{{\nu^{\prime}}^{2}-\nu^{2}}d\nu^{\prime}
\end{equation}
where $n_{0}$ is the value of $n$ at the high-frequency end of the infrared, and the integration is over the infrared region of the spectrum. In the case of pure CH$_{4}$, $n_{0}$ is equal to 1.32 at 15800~cm$^{-1}$ \citep{Trotta1996}. We will adopt the same value also for CH$_{4}$-N$_{2}$ mixtures.

After computing $n(\nu)$ and $k(\nu)$, we refine these calculations by taking into account first-order differences in transmission through boundaries between cell windows and vacuum in the case of empty cell, $T_{empty}(\nu)$, and between cell windows and ice in the case of filled cell, $T_{filled}(\nu)$. If the spectra are acquired by putting the beam on the side of the cell containing the KBr or CaF$_{2}$, we have to take into account the interfaces between the ice or vacuum and the KBr or CaF$_{2}$, as well (Figure \ref{cell}). In this case, we have:
\begin{equation}
k(\nu) = \frac{1}{4\pi\nu d}\Big[\ln \frac{T_{Al_{2}O_{3},ice} \: T_{ice,KBr}\:T_{KBr,ice}\:T_{ice,Al_{2}O_{3}}}{T_{Al_{2}O_{3},0}\:T_{0,KBr}\:T_{KBr,0}\:T_{0,Al_{2}O_{3}}}-\ln T(\nu)\Big] 
\end{equation}
where $T_{pq}$ represents the transmission across the boundary between the media with refractive index $m_{p}$ ($m_{p}=n_{p}+ik_{p}$) and $m_{q}$. Multiple reflections that are set up within the slab as the wave bounces
back and forth at the top and bottom interfaces are considered negligible. For the case of normal incidence we have:
\begin{equation}
T_{pq} = \frac{4\:n_{pq}}{(n_{pq}+1)^{2}+k_{pq}^{2}}
\end{equation}
where $n_{pq}$ and $k_{pq}$ are given by:
\begin{equation}
n_{pq} = \frac{(n_{p}\:n_{q}+k_{p}\:k_{q})}{n_{p}^2+k_{p}^{2}}
\end{equation}
\begin{equation}
k_{pq} = \frac{(n_{p}\:k_{q}-n_{q}\:k_{p})}{n_{p}^2+k_{p}^{2}}.
\end{equation}
The transmission spectrum and optical constants ($n$ and $k$) of pure CH$_{4}$ ice at 40 K are shown in four different bands in Figure~\ref{OpticalConstants}. 
\subsection{Pure CH$_{4}$ ice}\label{Temperature Effects}
\begin{figure}
	\ifincludegraphics
 	\centering
	\includegraphics[width=0.45\textwidth]{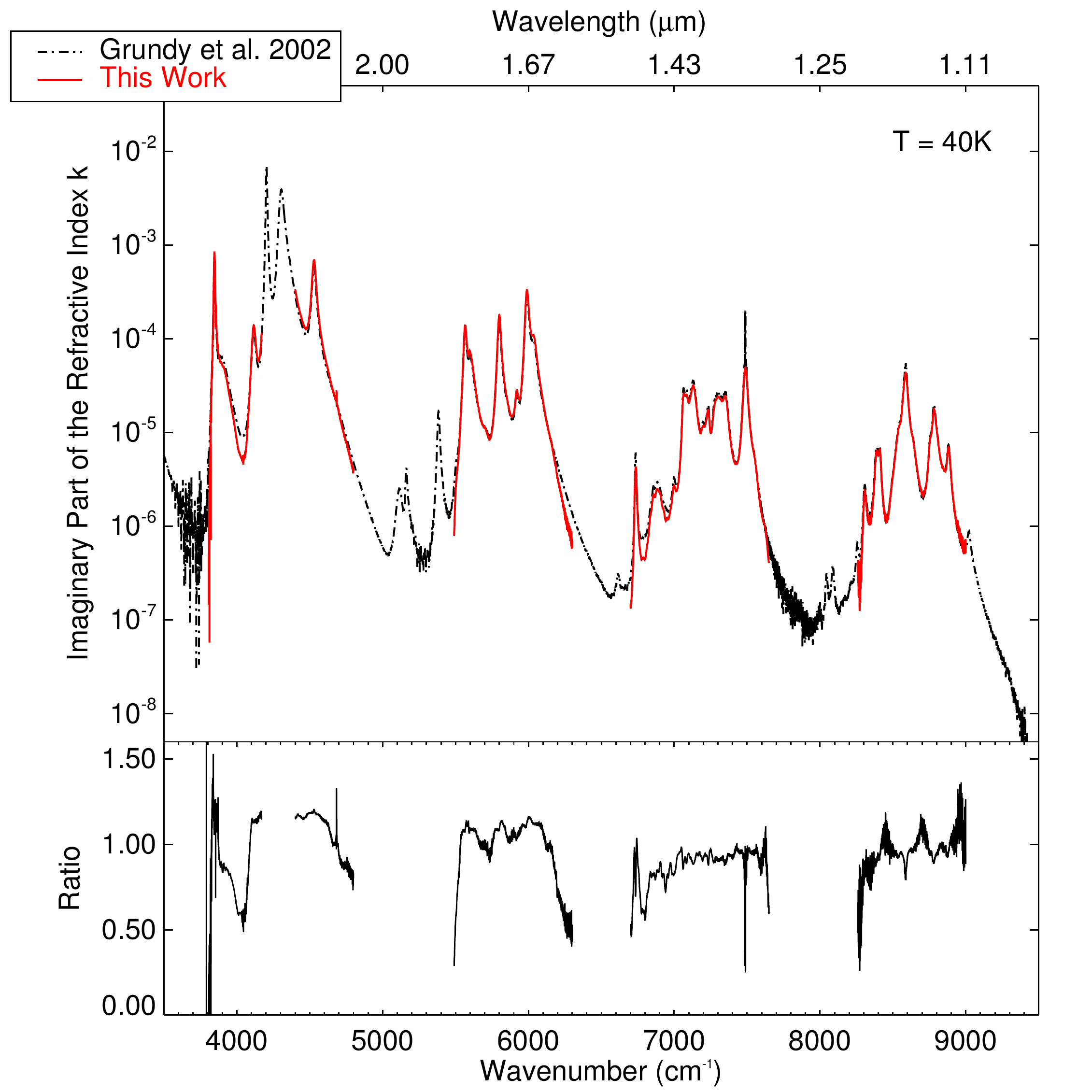}
	\fi
	\caption{Comparison between the imaginary part of the refractive index of pure CH$_{4}$ ice presented in this work (red solid line) and by \citet{Grundy2002} (black dash dot line) at 40~K. The ratio between the two data sets is shown in the bottom panel.}
	\label{ComparisonGrundy}%
\end{figure}

\citet{Grundy2002} present an exhaustive analysis of the temperature-dependent near-infrared absorption spectra of pure CH$_{4}$ ice, available at temperatures between 30 and 93~K. Our measurements for pure CH$_{4}$ ice at 40~K are compared with that by \citet{Grundy2002} in Figure \ref{ComparisonGrundy}. The absorption spectra, $\alpha$, by \citet{Grundy2002} have been converted for comparison purposes into the imaginary part of the refractive index, $k$, by means of the dispersion relation $\alpha=4 \pi k/\lambda$. While the data by \citet{Grundy2002} are obtained combining measurements acquired with different cell thicknesses, enabling measurement of the weakest and strongest CH$_{4}$ absorptions,  our data for pure \methane~ice correspond to a single sample of thickness 0.0252~cm. This prevents us from measuring absorption where it approaches zero or 100\%. Therefore, we provide our results in narrow blocks of wavelengths covering regions of intermediate absorption. Differences in band strength between our measurements and those by \citet{Grundy2002} are observed, as shown in Figure~\ref{ComparisonGrundy} for T= 40~K. Similar discrepancies between the two data sets are observed at all temperatures. This mismatch could be due to continuum removal, uncertainty in the sample thickness and/or temperature, transmission across the boundaries not taken into account when converting the absorption coefficients reported by \citet{Grundy2002} into $k$.
\begin{figure}[!h]
	\ifincludegraphics
 	\centering
	\includegraphics[width=0.5\textwidth]{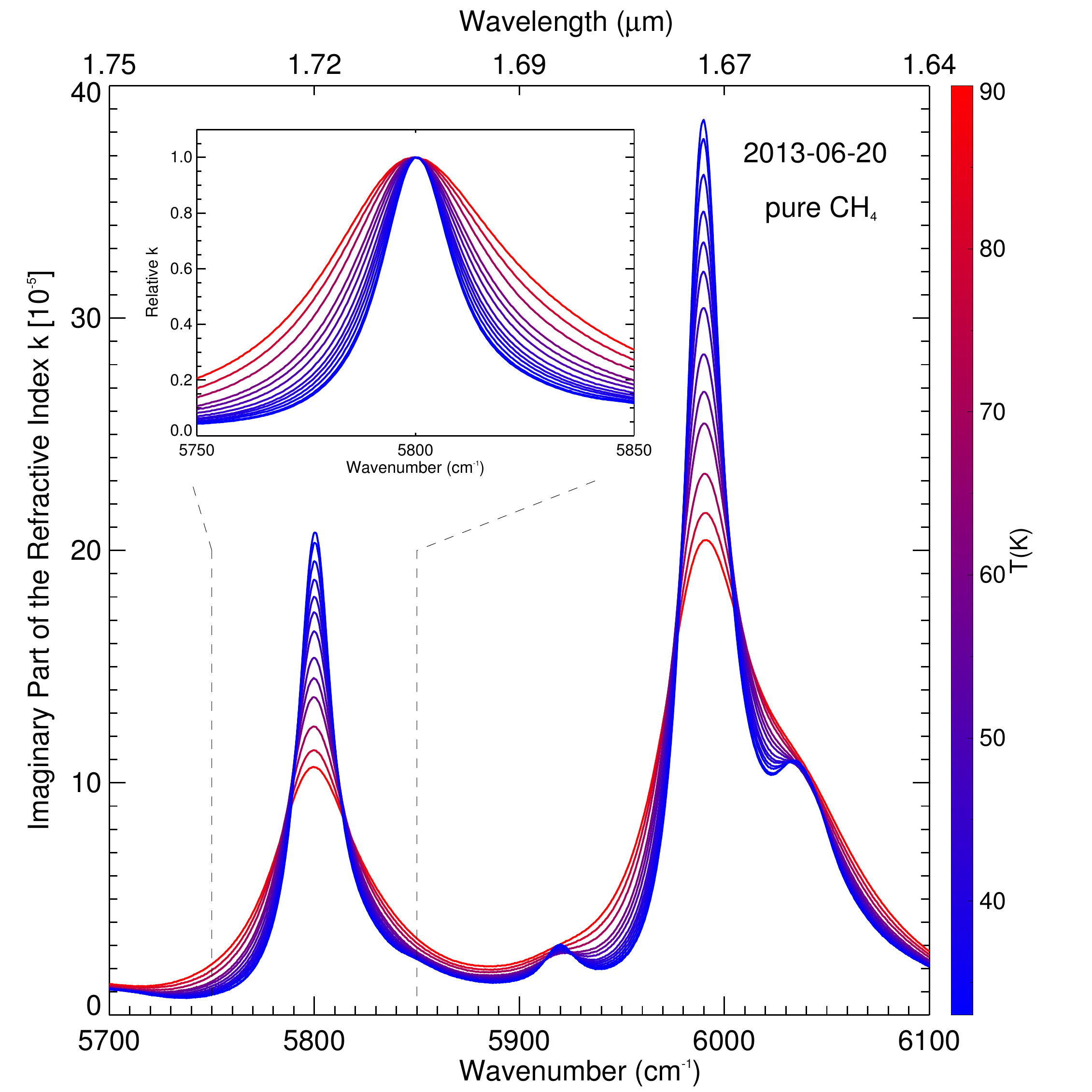}
	\fi
	\caption{$k$ spectra of pure CH$_{4}$ ice are shown for temperatures between 33 and 90~K. The inset panel is an enlargement of the region around 5800~cm$^{-1}$. In order to show, in greater detail, the thermal broadening of the $k$ spectra of pure CH$_{4}$ ice with temperature, the latter have been normalized to their maximum value in the range from 5750~cm$^{-1}$ to 5850~cm$^{-1}$.}
	\label{temperature_effects}%
\end{figure}
\begin{figure}[!h]
	\ifincludegraphics
 	\centering
	\includegraphics[width=0.5\textwidth]{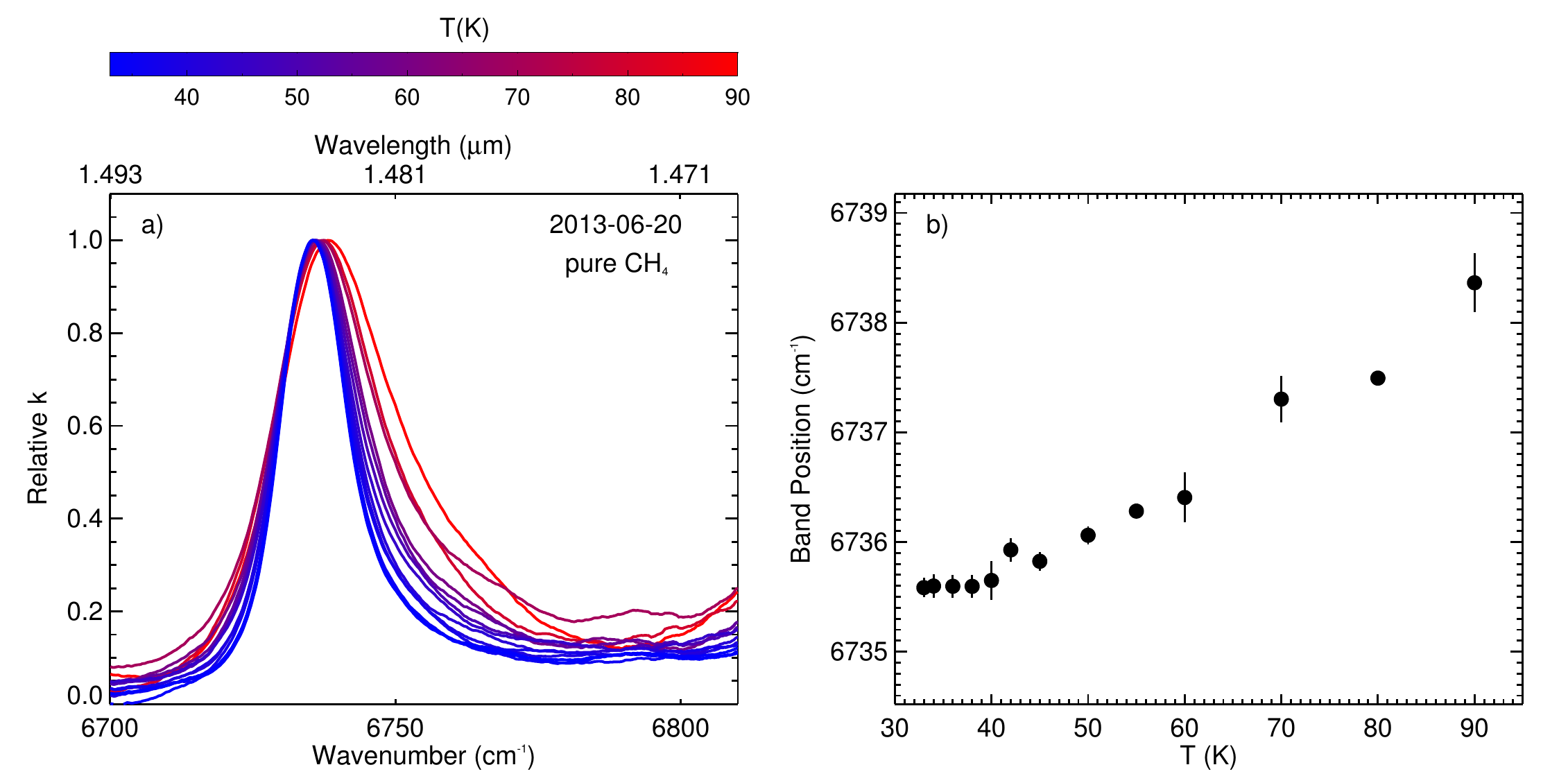}
	\fi
	\caption{Panel (a) shows $k$ spectra of pure CH$_{4}$ ice at temperatures between 33 and 90~K in the range between 6700 and 6810~cm$^{-1}$. The spectra have been normalized to their maximum value. The peak position as a function of temperature for the $\nu_{1}$+3$\nu_{4}$ CH$_{4}$ ice band is shown in panel (b).}
	\label{temperature_effects_1}%
\end{figure}

The general character of the temperature dependence of the pure CH$_{4}$ ice absorption coefficients discussed by \citet{Grundy2002} is confirmed by our data (Figure~\ref{temperature_effects}): lower temperature spectra exhibit narrower and higher peak absorptions which broaden and decrease in strength as the ice is warmed. Most of the CH$_{4}$ ice bands do not show significant temperature-dependent wavelength shifts. Nevertheless subtle shifts occasionally arise; a good example can be seen at 6735~cm$^{-1}$.  Figure~\ref{temperature_effects_1} displays the peak position of the $\nu_{1}$+3$\nu_{4}$ CH$_{4}$ ice band centered around 6735~cm$^{-1}$ as a function of temperature. The peak position has been estimated as the average of the position where the maximum of the band occurs and the maximum of the gaussian function fit to the data around the band center. The error is the dispersion between the two estimates. 
\subsection{CH$_{4}$-N$_{2}$ Solid Mixtures}    
The methane and nitrogen stoichiometry of a CH$_{4}$-N$_{2}$ solid mixture is estimated by dividing the integrated $k$ spectrum of the mixture by the corresponding integrated $k$ spectrum of pure CH$_{4}$ ice. This results in the fraction of CH$_{4}$ in the mixture. Given the temperature dependence of the pure CH$_{4}$ ice optical constants (see Section~\ref{Temperature Effects}), $k$ spectra  of CH$_{4}$-N$_{2}$ mixtures and pure CH$_{4}$ acquired at the same temperature are compared. This procedure is applied in the frequency ranges from 5300 -- 6440~cm$^{-1}$, 6700 -- 7650~cm$^{-1}$, and 8100 -- 9000~cm$^{-1}$. $F_{CH_{4},\nu}$ in Table \ref{tabA} represents the CH$_{4}$ fraction computed in the interval centered around $\nu$. The values obtained at all temperatures and frequency ranges for a given CH$_{4}$-N$_{2}$ mixture are averaged, resulting in the \methane~stoichiometry of our sample ($\overline{CH_{4}}$ [\%]). The mean CH$_{4}$ content, $\overline{CH_{4}}$ [\%], is listed in Table \ref{tabA} for all our experiments. The error on the $\overline{CH_{4}}$ concentration is given by the scatter of the values computed at all temperatures and frequency ranges and it is generally less than 5\%.  
\begin{figure}[!h]
	\ifincludegraphics
 	\centering
	\includegraphics[width=0.4\textwidth]{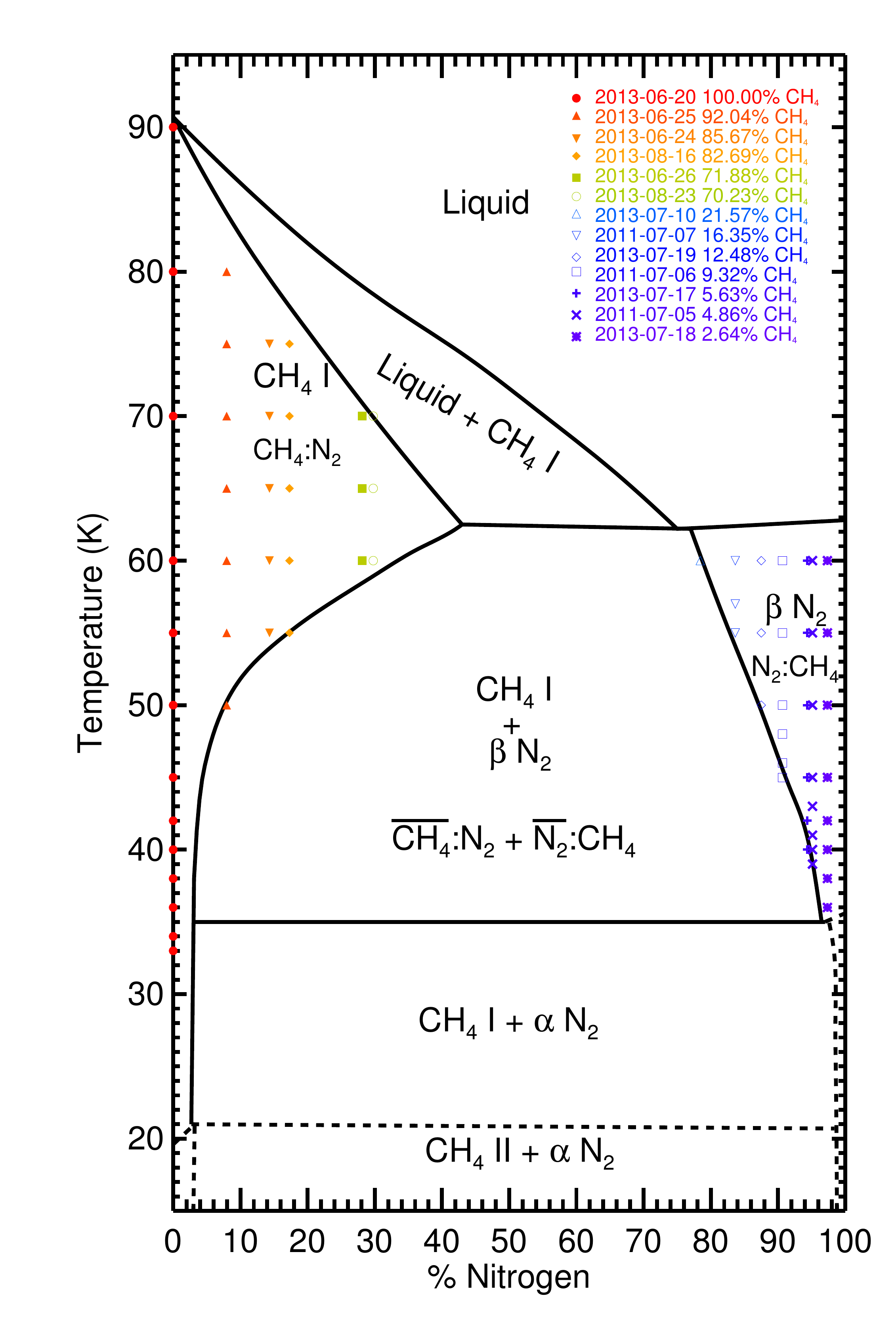}
	\fi
	\caption{The \methane-\nitrogen~binary phase diagram generated from X-ray diffraction studies by \citet{Prokhvatilov1983} is shown. The symbols indicate the different set of transmission measurements performed to conduct a systematic study of the changes in \methane-\nitrogen~solid mixtures
spectral behavior with mixing ratio and temperature. Measurements of \CHN~and \NCH~solid solutions acquired at different temperatures are represented by the same color and symbol. Between the solubility limits we have a mixture of two solid solution phases: \methane~ice saturated with \nitrogen, \phaseCHN, and \nitrogen~ice saturated with \methane, \phaseNCH.}
	\label{phase_diagram}%
\end{figure}
The measurements we present encompass both sides of the \methane-\nitrogen~phase diagram of \citet[][Figure~\ref{phase_diagram}]{Prokhvatilov1983} enabling us to characterize the spectral behavior of solid solutions of \methane~diluted
in \nitrogen~(\NCH) and \nitrogen~diluted in \methane~(\CHN) at different temperatures and mixing ratios. Example optical constants of \CHN~(92.04\% CH$_{4}$) and \NCH~(12.48\% CH$_{4}$) at two different temperatures at low spectral resolution are reported in Table \ref{OpticalConstantsTable}.

\begin{figure}[!h]
	\ifincludegraphics
 	\centering
	\includegraphics[width=\textwidth]{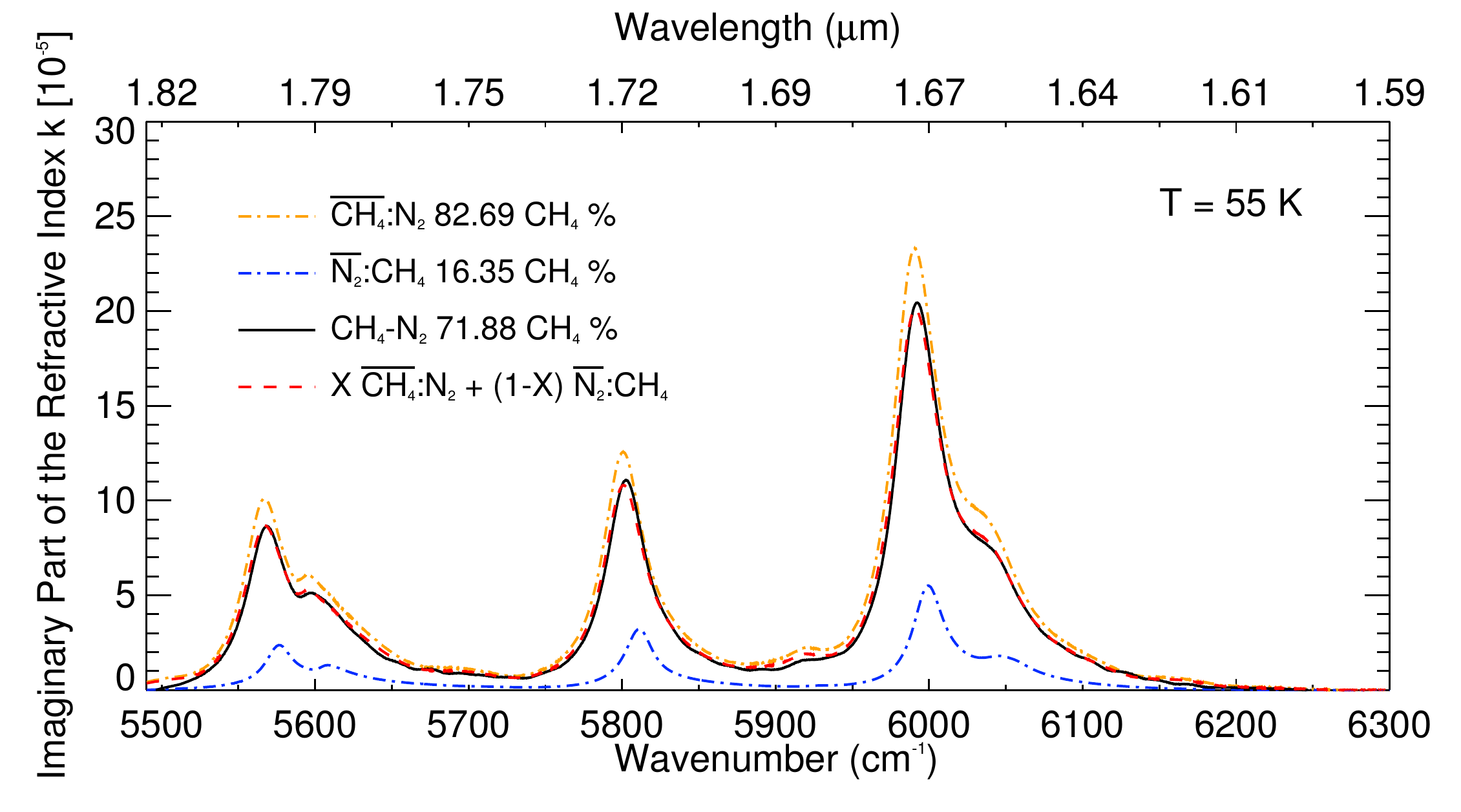}
	\fi
	\caption{$k$ spectrum of a  \methane-\nitrogen~mixture measured in the laboratory at 55~K and containing 71.88\% of \methane~(black solid line) and the synthetic $k$-spectrum (dashed red line) generated applying the lever rule to the optical constants of \phaseCHN~(dash-dot orange line) and \phaseNCH (dash-dot blue line) phases at 55~K. The synthetic spectrum reproduces the measured one well. }
	\label{leverrule}%
\end{figure}

Laboratory data of \CHN~and \NCH~at the solubility limits of \methane~and \nitrogen~in each other, which we indicate with \phaseCHN~and \phaseNCH~respectively, give the means to generate optical constants of \methane-\nitrogen~mixtures with any bulk composition between the solubility limits, at a given temperature (Figure~\ref{phase_diagram}). This is possible through the lever rule, which is a tool used to determine the molar percentage of each phase (in our case \phaseCHN~and \phaseNCH) in a two phase alloy in equilibrium. The molar percentage of the \phaseCHN~phase in a \methane-\nitrogen~mixture is
\begin{equation}
X = \frac{X_{CH_{4}}-S^{CH_{4}}_{\bf{\overline{N_{2}}}:CH_{4}}}{S^{CH_{4}}_{\bf{\overline{CH_{4}}}:N_{2}}-S^{CH_{4}}_{\bf{\overline{N_{2}}}:CH_{4}}},
\end{equation}
where $S^{CH_{4}}_{\bf{\overline{CH_{4}}}:N_{2}}$ and $S^{CH_{4}}_{\bf{\overline{N_{2}}}:CH_{4}}$ represent the solubility limit of \methane~in the \phaseCHN~and \phaseNCH~phases, respectively, and $X_{CH_{4}}$ is the overall \methane~abundance in the mixture. At 55~K the solubility limits of \methane~in the \methane-rich and \nitrogen-rich saturated phases are close to 82.69\% and 16.35\%, corresponding to measurements of optical constants we made in the laboratory (filled diamond and open downward triangle in Figure~\ref{phase_diagram}). It is therefore possible to generate synthetically the optical constants of any \methane-\nitrogen~mixture at 55~K. To a first order approximation our data behave consistent with expectation from the lever rule (see Figure~\ref{leverrule}).

\begin{figure*}
	\ifincludegraphics
 	\centering
	\includegraphics[width=0.9\textwidth]{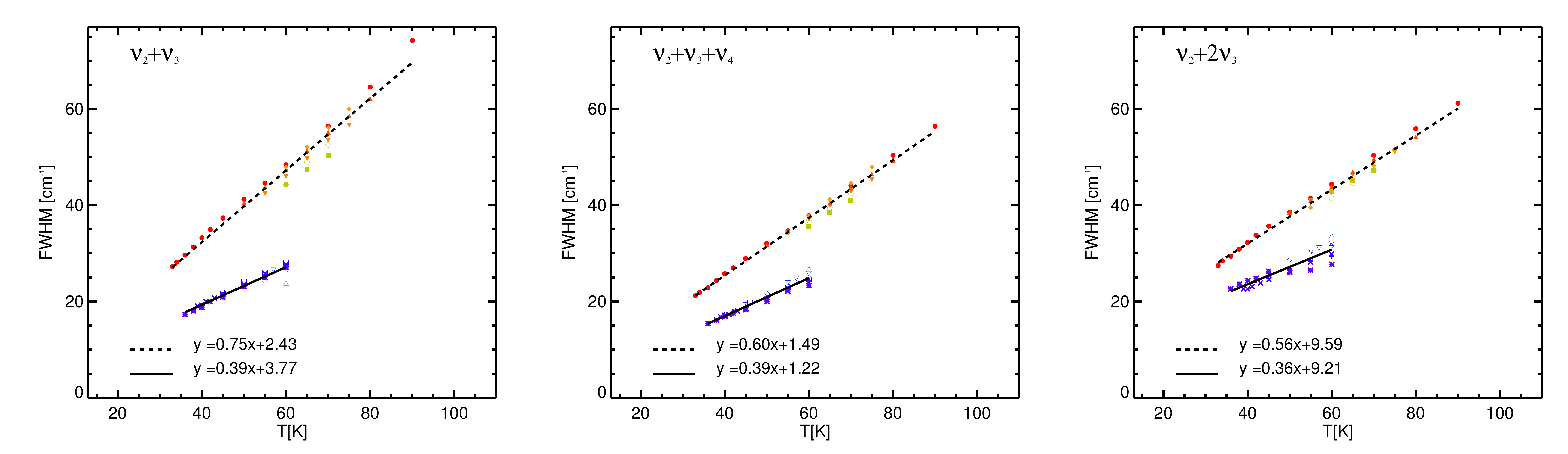}
	\fi
	\caption{Full width at half maximum (FWHM) as a function of temperature and stoichiometry for three isolated \methane-ice bands. Samples with different \methane~and \nitrogen~content are represented by different symbols (see legend in Figure~\ref{phase_diagram}). The dashed and solid lines represent the best fit of the \CHN~and \NCH~data, respectively.}
	\label{FWHM}%
\end{figure*}

Methane-nitrogen mixtures present a temperature-dependent behavior similar to that of pure \methane~(Section~\ref{Temperature Effects}). We characterize the thermal broadening of \methane~ice absorption bands by measuring their full widths at half maximum (FWHM) as a function of temperature and stoichiometry (Figure \ref{FWHM}). This analysis has been conducted for those \methane-bands that are isolated, not influenced by the wings of adjacent bands. These are the $\nu_{2}+\nu_{3}$, $\nu_{2}+\nu_{3}+\nu_{4}$, and $\nu_{2}+2\nu_{3}$ \methane-band, centered, in a sample of pure \methane~at 90~K, at 4526, 5800, and 7487~cm$^{-1}$, respectively. All bands we measured show a nearly linear dependence of FWHM with temperature, at all stoichiometries. Furthermore, the \methane~bands of \CHN~samples broaden faster than those of \NCH, as expected given the higher \methane~content in \CHN~than in \NCH~mixtures. 

\begin{figure}[!h]
	\ifincludegraphics
 	\centering
	\includegraphics[width=0.5\textwidth]{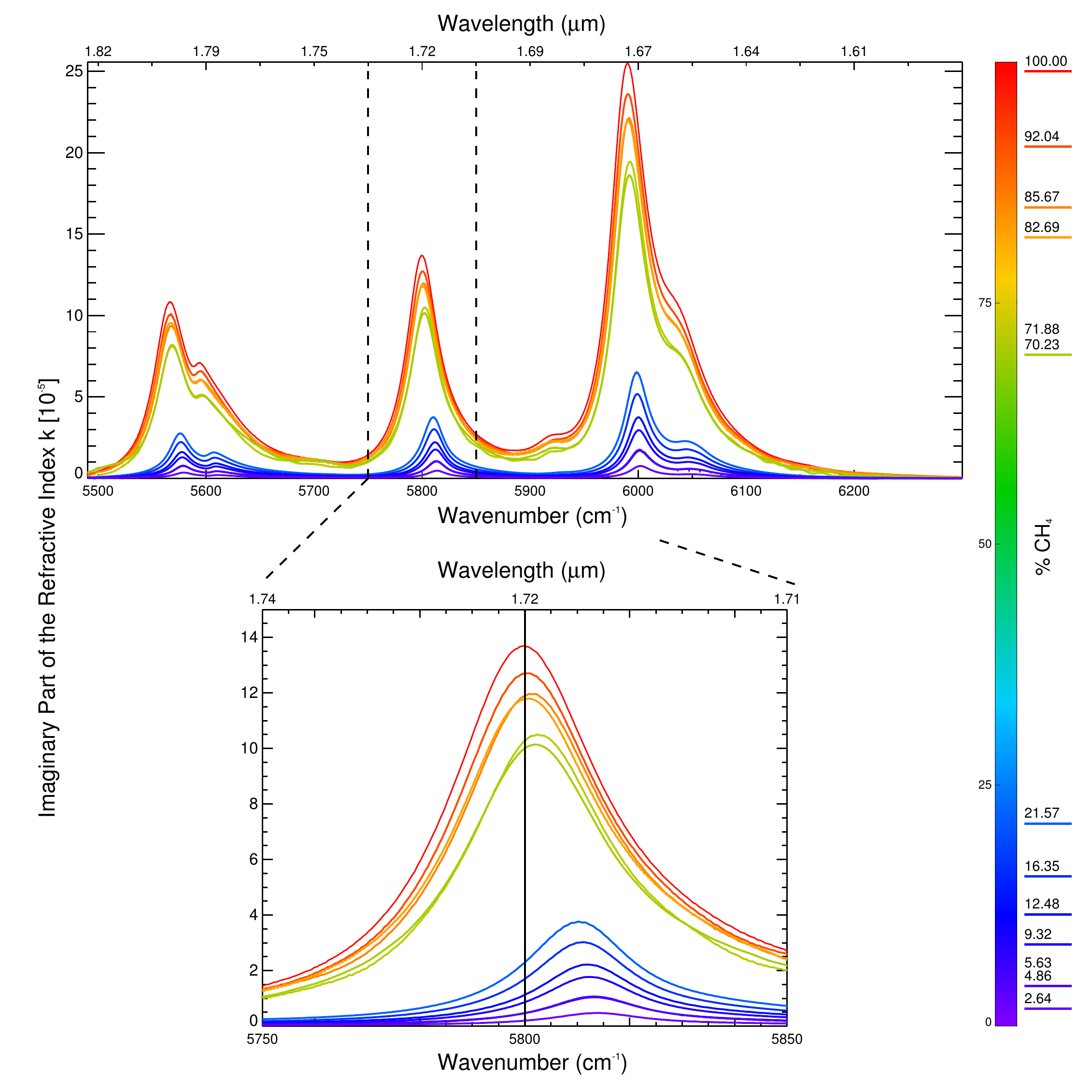}
	\fi
	\caption{\textit{Top panel}: The imaginary part of the refractive index, $k$, of CH$_{4}$-N$_{2}$ ice mixtures at different mixing ratios at T = 60 K 
over the range from 5490 to 6300 cm$^{-1}$. \textit{Bottom Panel}: Expanded view of the $\nu_{2}+\nu_{3}+\nu_{4}$ band.}
	\label{stoichiometry}%
\end{figure}

The imaginary part of the refractive index, $k$, of \methane-\nitrogen~ice mixtures at different mixing ratios at
T=60~K is shown in Figure \ref{stoichiometry}. \methane~when dissolved in \nitrogen~presents
absorption bands shifted toward shorter wavelengths compared
to the central wavelengths of pure \methane. This shift
varies with the \methane~abundance in the mixture: the larger the \methane~concentration
the smaller the blueshift. We estimate the spectral shift by means of a cross correlation technique. We shift a band in the pure \methane~spectrum relative to the same band in the mixture spectrum and for each shift a $\chi^{2}$ is computed. The $\chi^{2}$ versus shift behavior is fitted with a parabola, whose minimum position corresponds to the shift solution for the analyzed band. This method is described in detail by \citet{Tegler2008}. Another possible approach is, for a given band, to determine the difference between the peak positions in the \methane-\nitrogen~and pure \methane~samples, with the peak position estimated as described in Section \ref{Temperature Effects}\null. The solutions obtained with these different techniques are averaged, and the dispersion between them is taken to be representative of the uncertainty. In order to correctly determine the blueshift of the \methane-\nitrogen~bands relative to pure \methane, spectra acquired at similar temperatures are compared, given the temperature dependent behavior of the \methane~bands peak position (Section~\ref{Temperature Effects}).
\begin{figure}[!h]
	\ifincludegraphics
 	\centering
	\includegraphics[width=0.5\textwidth]{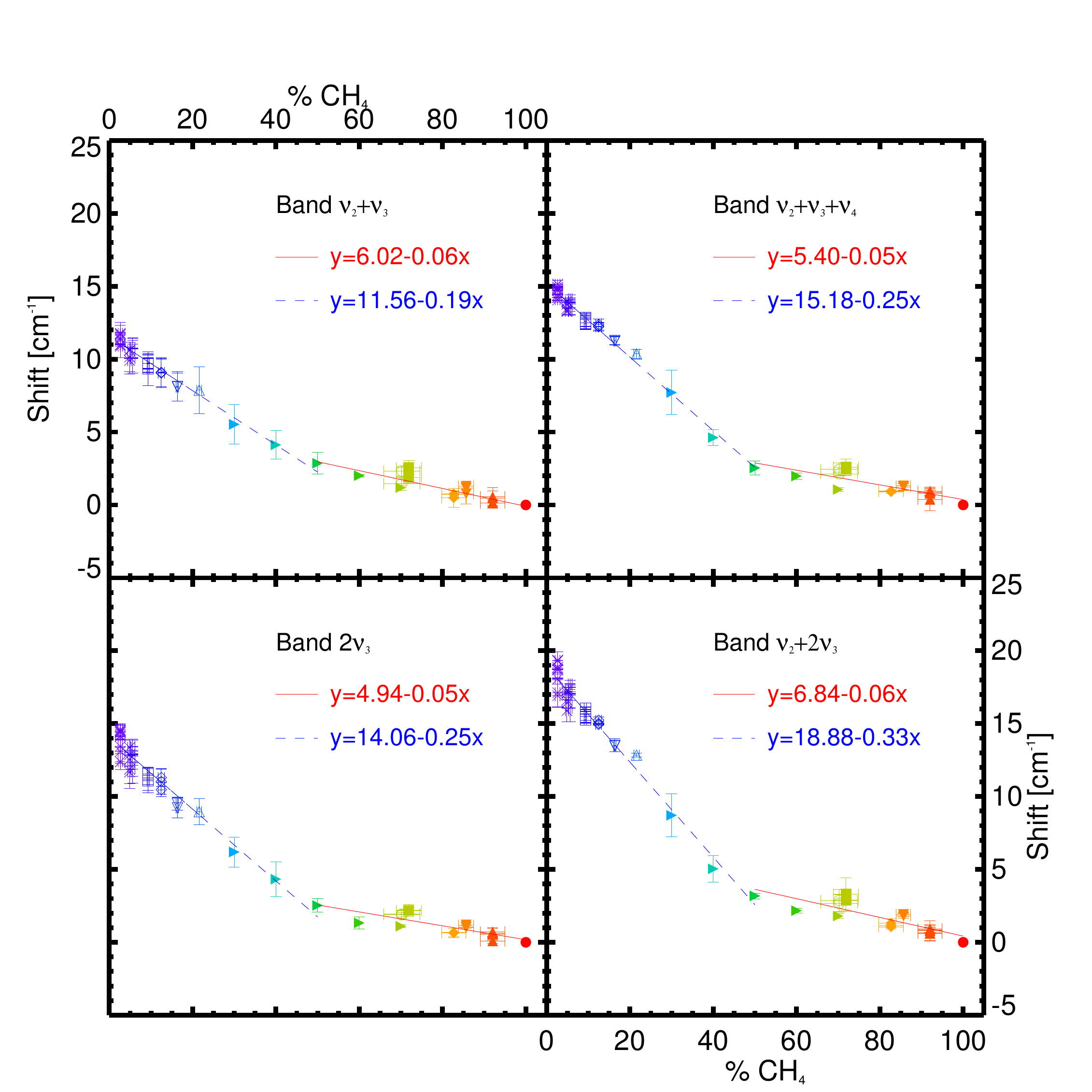}
	\fi
	\caption{Blueshifts of the $\nu_{2}+\nu_{3}$ (4526~cm$^{-1}$), $\nu_{2}+\nu_{3}+\nu_{4}$ (5800~cm$^{-1}$), $2\nu_{3}$ (5991~cm$^{-1}$), and $\nu_{2}+2\nu_{3}$ (7487~cm$^{-1}$) bands versus \methane~abundance. Symbols and colors correspond to samples with different \methane~and \nitrogen~stoichiometries as described by the legend in Figure \ref{phase_diagram} (see text for details). Solid and dashed lines are the linear fit to the shift versus \methane~abundance for percentages greater and less than 50\%, respectively.}
	\label{blueshift}%
\end{figure}
The blueshift of four \methane~bands versus \methane~abundance is reported in Figure~\ref{blueshift}. For a given \methane~stoichiometry, different measurements correspond to different temperatures. No significant variations in the wavelength shift with respect to temperature are observed. Filled rightfacing triangles in Figure~\ref{blueshift}, contrary to the other symbols that represent samples generated in the laboratory (see legend in Figure~\ref{phase_diagram} and Table~\ref{tabA}), refer to optical constants computed synthetically by applying the lever rule at 55~K for a range of \methane~abundance between 30\% and 70\%, in 10\% steps. 
The \nitrogen-enriched and \methane-enriched sides of the phase diagram present a different linear trend of the blueshift versus \methane~abundance (dashed and solid lines in Figure~\ref{phase_diagram}). Such trend varies from band to band. 
\section{Applications to Pluto}
Efforts to quantitatively model Pluto's spectrum over the course of the years identified several important issues. The first
concerns the state of \methane~ice on Pluto's surface. Seen
at high spectral resolution, the \methane~bands in Pluto's spectrum are
shifted toward shorter wavelengths compared to the central wavelengths of
pure \methane~obtained in the laboratory, implying \methane~being dissolved at low concentrations
in a matrix of solid \nitrogen~\citep{Schmitt1992,Quirico1997}. Additionally, the identification of the 1.69-\micron~band in Pluto's spectrum has been interpreted so far as the evidence of the presence of pure \methane~on Pluto. Indeed, the 1.69-\micron~\methane~feature has never been observed in any sample of
\methane~diluted at low concentrations in \nitrogen~(neither in $\alpha$- or $\beta$-N$_{2}$ phases)
at any temperature. These considerations together with the lack of optical constants for \methane~with a small fraction of \nitrogen~are the main arguments behind the approach taken in analyzing
Pluto spectra in the last decade \citep[\eg,][]{doute1999,olkin2007,Protopapa2008}, employing pure \methane~and \methane~diluted at low concentrations in \nitrogen. 

The set of optical constants presented in this work enable us to compare and contrast samples with \methane~dissolved at low concentrations
in a matrix of solid \nitrogen~(\NCH), and  \methane~with a small fraction of \nitrogen~(\CHN). These data reveal that the 1.69-$\mu$m band of Pluto's spectrum can no longer be considered as evidence for the presence of pure \methane~on Pluto's surface, as this feature is observed not only in samples of pure \methane~(at temperatures below $\sim$60~K, Figure~\ref{temperature_effects}), but also in \CHN~samples (Figure~\ref{stoichiometry}). Also, the wavelength shift of Pluto's \methane~bands with respect to that of pure \methane, does not necessarily imply the presence of \methane~being dissolved at low concentrations
in a matrix of solid \nitrogen, as the wavelength shift is a linear function of the \methane~abundance (Figure~\ref{blueshift}).

The quantitative analysis of the wavelength shift versus abundance indicates that the wavelength shift of the \methane~bands is indeed a good dilution indicator. The linear fits of the wavelength shift as a function of \methane~abundance for different \methane~bands provide the tools to infer the \methane-\nitrogen~mixing ratio on the surface of Pluto, or any TNO covered by \methane~and \nitrogen. Also, it is extremely important to stress that the linear dependency of the shift versus \methane~abundance changes from band to band. This needs to be taken into account when comparing the shift of deeper and shallower bands to address stratification properties \citep{Merlin2009}.

Our new optical constants enable us to model for the first time Pluto's near-infrared spectroscopic measurements in compliance with thermodynamic equilibrium, superseding more than a decade of work in which Pluto's spectrum was interpreted in terms of a combination of ``diluted'' and ``pure'' \methane~components. To account for thermodynamic equilibrium on Pluto's surface, we need instead to consider methane saturated with nitrogen, \phaseCHN, and nitrogen saturated with methane, \phaseNCH, as dictated by the \nitrogen-\methane~binary phase diagram \citep[Figure~\ref{phase_diagram},][]{Prokhvatilov1983} and suggested by \citet{Trafton2015} and \citet{Cruikshank2015}. Specifically, for a plausible Pluto volatile ice temperature of 40~K \citep{Tryka1994}, the diagram shows that the solubility limits of \methane~and \nitrogen~in each other are about 5\% (\phaseNCH, with 5\% \methane) and 3\% (\phaseCHN~with 3\% \nitrogen). While we have optical constants available for the former phase (see Figure \ref{phase_diagram}), the latter is not available yet. However, we have laboratory data covering a wide range of temperatures and mixing ratios to synthetically reproduce the lacking set of optical constants. The laboratory measurements presented in this work cover the wavelength range of the LEISA spectrometer on board of the New Horizons spacecraft \citep{Young2008}, which will flyby Pluto in July 2015 and therefore will help to improve the scientific outcome of this mission.

\section{Acknowledgments}

The authors gratefully thank NASA's Outer Planets Research program (grant
\#NNX11AM53G) for funding that supported this work and the Mt.\ Cuba
Astronomical Foundation for hardware upgrades used in the experiments
described in this paper.  J. Bergonio thanks the National Science
Foundation's Research Experience for Undergraduates grant AST-1004107
to Northern Arizona University that enabled his contribution.  We also
thank two anonymous reviewers for their constructive suggestions that
helped to improve this paper and M. Adamkovics for valuable pointers on
implementation of the Kramers-Kronig calculation.  Finally, we thank
the free and open source software communities for empowering us with
key software tools used to complete this project.





\bibliographystyle{icarus}
\bibliography{silvia}
\newpage
	\afterpage{
	\clearpage
	\begin{longtable} {c c c c c c c c}
	\caption{Stoichiometry}\label{tabA}\\
	\hline
	\hline
         	\textbf{Date} &
		\textbf{Spacer}	& 
		\textbf{$d$} & 
		\textbf{T} 	&
		\textbf{$F_{CH_{4},5870}$} &
		\textbf{$F_{CH_{4}, 7175}$} &	
		\textbf{$F_{CH_{4}, 8550}$} &
		\textbf{$\overline{CH_{4}}$} \\
		{YYYY-MM-DD}	 	&
					 		&
		{\textbf{[cm]}} 			& 
		{[K]}			 		&	
		{\%}			 		& 	
		{\%}					& 	
		{\%}	 				& 	
		{\%}					\\
	
	\hline
	2011-07-05 	& n/a							& 0.520		&	60	&	5.21				&	4.26				&	4.76				&\\
	2011-07-05	&n/a							& 0.520		&	55	&	5.27				&	4.26				&	4.82				&\\
	2011-07-05	&n/a							& 0.520		&	50	&	5.29				&	4.28				&	4.89				&\\
	2011-07-05	&n/a							& 0.520		&	45	&	5.23				&	4.61				&	5.17				&\\
	2011-07-05	&n/a							& 0.520		&	40	&	5.35				&	4.40				&	5.06				&\\
	\rowcolor{LightGoldenrodYellow} 
	2011-07-05	&n/a							& 0.520		&		&					&					&					&	4.86$\pm$0.41		\\
	2011-07-06	&n/a							& 0.520		&	60	&	8.92 				&	9.20				&	9.45				&\\
	2011-07-06	&n/a							& 0.520		&	55	&	9.00 				&	9.29				&	9.63				&\\
	2011-07-06	&n/a							& 0.520		&	50	&	9.08 				&	9.31				&	9.64				&\\
	2011-07-06	&n/a							& 0.520		&	45	&	9.13 				&	9.37				&	9.84				&\\
	\rowcolor{LightGoldenrodYellow} 2011-07-06        &n/a          	& 0.520		&		&					&					&					&	9.32$\pm$0.28		\\
	2011-07-07	&n/a							& 0.520		&	60	&	16.04 			&	16.21			&	16.54			&\\
	2011-07-07	&n/a							& 0.520		&	55	&	16.16 			&	16.33			&	16.80			&\\
	\rowcolor{LightGoldenrodYellow} 2011-07-07        &n/a          	& 0.520		&		&					&					&					&	16.35$\pm$0.28	\\
	2013-06-24	&KBr							& 0.025		&	70	&	85.08     			&	84.71			&	88.06			&\\
	2013-06-24	&KBr							& 0.025		&	60	&	85.89     			&	82.77			&	87.58			&\\
	2013-06-24	&KBr							& 0.025		&	55	&	86.43     			&	83.68			&	86.82			&\\	
	\rowcolor{LightGoldenrodYellow} 2013-06-24        & KBr        	& 0.025		&		&					&					&					&	85.67$\pm$1.77	\\
	2013-06-25	&KBr							& 0.025		&	80	&	91.30     			&	89.10			&	96.13			&\\
	2013-06-25	&KBr							& 0.025		&	70	&	90.86     			&	90.47			&	88.91			&\\
	2013-06-25	&KBr							& 0.025		&	60	&	91.73     			&	89.15			&	96.12			&\\
	2013-06-25	&KBr							& 0.025		&	55	&	91.74     			&	89.71			&	96.68			&\\
	2013-06-25	&KBr							& 0.025		&	50	&	91.31     			&	90.37			&	97.09			&\\
	\rowcolor{LightGoldenrodYellow} 2013-06-25        & KBr       		& 0.025		&		&					&					&					&	92.04$\pm$2.93	\\
	2013-06-26	&KBr							& 0.025		&	70	&	71.01     			&	68.00			&	75.16			&\\
	2013-06-26	&KBr							& 0.025		&	60	&	71.28     			&	70.05			&	75.76			&\\
	\rowcolor{LightGoldenrodYellow} 2013-06-26        & KBr       		& 0.025		&		&					&					&					&	71.88$\pm$3.01	\\
	2013-07-10        & n/a       						& 0.522		&	60	&	20.89     			&	21.98			&	21.84			&\\
	\rowcolor{LightGoldenrodYellow} 2013-07-10        & n/a       		& 0.522		&		&					&					&					&	21.57$\pm$0.59	\\
	2013-07-17        & n/a       						& 0.522		&	60	&	5.46     			&	5.76				&	4.51				&\\
	2013-07-17        & n/a       						& 0.522		&	55	&	5.28     			&	5.80				&	5.48				&\\
	2013-07-17        & n/a       						& 0.522		&	50	&	5.26     			&	5.74				&	5.59				&\\
	2013-07-17        & n/a       						& 0.522		&	45	&	5.38     			&	5.91				&	5.38				&\\
	2013-07-17        & n/a       						& 0.522		&	42	&	5.32     			&	5.93				&	5.82				&\\
	2013-07-17        & n/a       						& 0.522		&	40	&	5.45     			&	6.61				&	6.68				&\\
	\rowcolor{LightGoldenrodYellow} 2013-07-17        	& n/a       		& 0.522		&		&					&					&					&	5.63 $\pm$0.49	\\
	2013-07-18        & n/a       						& 0.522		&	60	&	2.47     			&	2.05				&	1.80				&\\
	2013-07-18        & n/a       						& 0.522		&	55	&	2.51     			&	2.10				&	1.55				&\\
	2013-07-18        & n/a       						& 0.522		&	50	&	2.33     			&	2.26				&	2.47				&\\
	2013-07-18        & n/a       						& 0.522		&	45	&	2.23     			&	2.70				&	3.60				&\\
	2013-07-18        & n/a       						& 0.522		&	42	&	2.46     			&	2.92				&	3.06				&\\
	2013-07-18        & n/a       						& 0.522		&	40	&	2.41     			&	3.02				&	3.56				&\\
	2013-07-18        & n/a       						& 0.522		&	38	&	2.43     			&	3.15				&	3.63				&\\
	2013-07-18        & n/a       						& 0.522		&	36	&	2.45     			&	3.14				&	3.17				&\\
	\rowcolor{LightGoldenrodYellow} 2013-07-18        	& n/a       		& 0.522		&		&					&					&					&	2.64 $\pm$0.56	\\
	2013-07-19        & n/a       						& 0.522		&	60	&	11.62     			&	12.44			&	12.25			&\\
	2013-07-19        & n/a       						& 0.522		&	55	&	11.63     			&	12.77			&	13.24			&\\
	2013-07-19        & n/a       						& 0.522		&	50	&	11.79     			&	13.16			&	13.46			&\\
	\rowcolor{LightGoldenrodYellow} 2013-07-19        & n/a       		& 0.522		&		&					&					&					&	12.48 $\pm$0.71	\\
	2013-08-16        & CaF$_2$  					& 0.011		&	70	&	85.64     			&	78.96			&	83.40			&\\
	2013-08-16        & CaF$_2$  					& 0.011		&	60	&	85.84     			&	78.15			&	82.11			&\\
	2013-08-16        & CaF$_2$  					& 0.011		&	55	&	85.25     			&	80.55			&	84.31			&\\
	\rowcolor{LightGoldenrodYellow} 2013-08-16        & CaF$_2$	& 0.011		&		&					&					&					&	82.69 $\pm$2.91	\\
	2013-08-23        & CaF$_2$  					& 0.011		&	70	&	70.27     			&	68.48			&	69.71			&\\
	2013-08-23        & CaF$_2$  					& 0.011		&	60	&	72.36     			&	63.67			&	76.89			&\\
	\rowcolor{LightGoldenrodYellow} 2013-08-23        & CaF$_2$	& 0.011		&		&					&					&					&	70.23 $\pm$4.36	\\
	\hline	
	\hline 	
	\end{longtable}
	}
	
\newpage	
	\afterpage{
	\clearpage
	\begin{table}[!b]
	\tiny
	\caption{Optical constants of two different CH$_{4}$-N$_{2}$ mixtures at two different temperatures in the range from 5500 to 5900 cm$^{-1}$.}
	\begin{center}
	\resizebox {\textwidth }{!}{%
	\begin{tabular}{clrclrclrclr}
	\hline
	& \multicolumn{5}{c}{92.04\% CH$_{4}$ in N$_{2}$} & &\multicolumn{5}{c}{12.48\% CH$_{4}$ in N$_{2}$}\\
	& \multicolumn{2}{c}{T = 60 K} & &\multicolumn{2}{c}{T = 50 K} & &  \multicolumn{2}{c}{T = 60 K} & &\multicolumn{2}{c}{T = 50 K}\\
	\cline{2-3} \cline{5-6} \cline{8-9} \cline{11-12}
	Wavenumber (cm$^{-1}$) & n & k & & n & k & &n & k & & n & k\\
    5500.08  & 1.3198817  & 0.0000040  &  & 1.3198855  & 0.0000033  &  & 1.3199970  & 0.0000004  &  & 1.3199971  & 0.0000002 \\
   5507.35  & 1.3198864  & 0.0000054  &  & 1.3198900  & 0.0000044  &  & 1.3199974  & 0.0000006  &  & 1.3199975  & 0.0000003 \\
   5514.62  & 1.3198919  & 0.0000077  &  & 1.3198955  & 0.0000059  &  & 1.3199978  & 0.0000008  &  & 1.3199980  & 0.0000005 \\
   5521.89  & 1.3198977  & 0.0000111  &  & 1.3199019  & 0.0000088  &  & 1.3199983  & 0.0000010  &  & 1.3199986  & 0.0000006 \\
   5529.16  & 1.3199049  & 0.0000158  &  & 1.3199099  & 0.0000129  &  & 1.3199989  & 0.0000013  &  & 1.3199992  & 0.0000009 \\
   5536.43  & 1.3199121  & 0.0000237  &  & 1.3199179  & 0.0000204  &  & 1.3199997  & 0.0000017  &  & 1.3200001  & 0.0000012 \\
   5543.71  & 1.3199202  & 0.0000348  &  & 1.3199281  & 0.0000308  &  & 1.3200006  & 0.0000022  &  & 1.3200011  & 0.0000018 \\
   5550.98  & 1.3199250  & 0.0000542  &  & 1.3199366  & 0.0000514  &  & 1.3200017  & 0.0000032  &  & 1.3200023  & 0.0000027 \\
   5558.25  & 1.3199178  & 0.0000812  &  & 1.3199321  & 0.0000846  &  & 1.3200030  & 0.0000048  &  & 1.3200039  & 0.0000044 \\
   5565.52  & 1.3198898  & 0.0001000  &  & 1.3198977  & 0.0001113  &  & 1.3200043  & 0.0000081  &  & 1.3200057  & 0.0000078 \\
   5572.79  & 1.3198601  & 0.0000923  &  & 1.3198591  & 0.0000989  &  & 1.3200024  & 0.0000139  &  & 1.3200045  & 0.0000149 \\
   5580.06  & 1.3198488  & 0.0000746  &  & 1.3198484  & 0.0000755  &  & 1.3199948  & 0.0000155  &  & 1.3199947  & 0.0000175 \\
   5587.33  & 1.3198508  & 0.0000636  &  & 1.3198535  & 0.0000623  &  & 1.3199916  & 0.0000109  &  & 1.3199911  & 0.0000113 \\
   5594.60  & 1.3198495  & 0.0000659  &  & 1.3198546  & 0.0000666  &  & 1.3199922  & 0.0000083  &  & 1.3199921  & 0.0000083 \\
   5601.88  & 1.3198413  & 0.0000604  &  & 1.3198454  & 0.0000625  &  & 1.3199935  & 0.0000078  &  & 1.3199937  & 0.0000078 \\
   5609.15  & 1.3198380  & 0.0000527  &  & 1.3198408  & 0.0000547  &  & 1.3199925  & 0.0000093  &  & 1.3199928  & 0.0000098 \\
   5616.42  & 1.3198366  & 0.0000456  &  & 1.3198388  & 0.0000467  &  & 1.3199908  & 0.0000081  &  & 1.3199908  & 0.0000086 \\
   5623.69  & 1.3198371  & 0.0000391  &  & 1.3198392  & 0.0000392  &  & 1.3199903  & 0.0000066  &  & 1.3199902  & 0.0000068 \\
   5630.96  & 1.3198383  & 0.0000327  &  & 1.3198408  & 0.0000327  &  & 1.3199903  & 0.0000054  &  & 1.3199902  & 0.0000055 \\
   5638.23  & 1.3198406  & 0.0000269  &  & 1.3198434  & 0.0000267  &  & 1.3199906  & 0.0000045  &  & 1.3199906  & 0.0000045 \\
   5645.50  & 1.3198447  & 0.0000224  &  & 1.3198473  & 0.0000220  &  & 1.3199910  & 0.0000038  &  & 1.3199910  & 0.0000038 \\
   5652.78  & 1.3198489  & 0.0000194  &  & 1.3198518  & 0.0000187  &  & 1.3199914  & 0.0000033  &  & 1.3199914  & 0.0000032 \\
   5660.05  & 1.3198527  & 0.0000174  &  & 1.3198558  & 0.0000167  &  & 1.3199918  & 0.0000029  &  & 1.3199918  & 0.0000028 \\
   5667.32  & 1.3198560  & 0.0000159  &  & 1.3198594  & 0.0000152  &  & 1.3199922  & 0.0000025  &  & 1.3199922  & 0.0000025 \\
   5674.59  & 1.3198593  & 0.0000146  &  & 1.3198626  & 0.0000138  &  & 1.3199925  & 0.0000022  &  & 1.3199925  & 0.0000022 \\
   5681.86  & 1.3198625  & 0.0000133  &  & 1.3198659  & 0.0000124  &  & 1.3199929  & 0.0000020  &  & 1.3199930  & 0.0000019 \\
   5689.13  & 1.3198659  & 0.0000120  &  & 1.3198696  & 0.0000115  &  & 1.3199933  & 0.0000018  &  & 1.3199934  & 0.0000017 \\
   5696.40  & 1.3198695  & 0.0000115  &  & 1.3198730  & 0.0000112  &  & 1.3199937  & 0.0000017  &  & 1.3199939  & 0.0000016 \\
   5703.67  & 1.3198729  & 0.0000109  &  & 1.3198761  & 0.0000106  &  & 1.3199942  & 0.0000015  &  & 1.3199943  & 0.0000015 \\
   5710.95  & 1.3198766  & 0.0000102  &  & 1.3198797  & 0.0000098  &  & 1.3199944  & 0.0000014  &  & 1.3199947  & 0.0000014 \\
   5718.22  & 1.3198805  & 0.0000097  &  & 1.3198832  & 0.0000092  &  & 1.3199949  & 0.0000013  &  & 1.3199950  & 0.0000013 \\
   5725.49  & 1.3198850  & 0.0000092  &  & 1.3198878  & 0.0000085  &  & 1.3199954  & 0.0000013  &  & 1.3199955  & 0.0000012 \\
   5732.76  & 1.3198906  & 0.0000092  &  & 1.3198932  & 0.0000081  &  & 1.3199960  & 0.0000012  &  & 1.3199961  & 0.0000010 \\
   5740.03  & 1.3198967  & 0.0000108  &  & 1.3198994  & 0.0000092  &  & 1.3199965  & 0.0000012  &  & 1.3199966  & 0.0000010 \\
   5747.30  & 1.3199028  & 0.0000125  &  & 1.3199056  & 0.0000106  &  & 1.3199972  & 0.0000013  &  & 1.3199973  & 0.0000011 \\
   5754.57  & 1.3199103  & 0.0000153  &  & 1.3199134  & 0.0000128  &  & 1.3199980  & 0.0000014  &  & 1.3199981  & 0.0000012 \\
   5761.85  & 1.3199188  & 0.0000206  &  & 1.3199224  & 0.0000172  &  & 1.3199987  & 0.0000016  &  & 1.3199990  & 0.0000014 \\
   5769.12  & 1.3199279  & 0.0000286  &  & 1.3199327  & 0.0000245  &  & 1.3199998  & 0.0000020  &  & 1.3200001  & 0.0000017 \\
   5776.39  & 1.3199381  & 0.0000422  &  & 1.3199451  & 0.0000375  &  & 1.3200010  & 0.0000027  &  & 1.3200012  & 0.0000023 \\
   5783.66  & 1.3199441  & 0.0000655  &  & 1.3199555  & 0.0000617  &  & 1.3200024  & 0.0000038  &  & 1.3200029  & 0.0000033 \\
   5790.93  & 1.3199365  & 0.0000984  &  & 1.3199521  & 0.0001012  &  & 1.3200042  & 0.0000058  &  & 1.3200051  & 0.0000052 \\
   5798.20  & 1.3199021  & 0.0001253  &  & 1.3199118  & 0.0001395  &  & 1.3200057  & 0.0000098  &  & 1.3200072  & 0.0000093 \\
   5805.47  & 1.3198574  & 0.0001181  &  & 1.3198533  & 0.0001296  &  & 1.3200043  & 0.0000172  &  & 1.3200068  & 0.0000179 \\
   5812.74  & 1.3198355  & 0.0000899  &  & 1.3198298  & 0.0000914  &  & 1.3199944  & 0.0000220  &  & 1.3199950  & 0.0000252 \\
   5820.02  & 1.3198333  & 0.0000656  &  & 1.3198310  & 0.0000633  &  & 1.3199866  & 0.0000155  &  & 1.3199850  & 0.0000166 \\
   5827.29  & 1.3198375  & 0.0000499  &  & 1.3198379  & 0.0000468  &  & 1.3199866  & 0.0000096  &  & 1.3199856  & 0.0000095 \\
   5834.56  & 1.3198431  & 0.0000387  &  & 1.3198450  & 0.0000361  &  & 1.3199880  & 0.0000068  &  & 1.3199877  & 0.0000065 \\
   5841.83  & 1.3198498  & 0.0000310  &  & 1.3198527  & 0.0000289  &  & 1.3199890  & 0.0000052  &  & 1.3199888  & 0.0000050 \\
   5849.10  & 1.3198564  & 0.0000264  &  & 1.3198596  & 0.0000249  &  & 1.3199899  & 0.0000043  &  & 1.3199899  & 0.0000040 \\
   5856.37  & 1.3198618  & 0.0000227  &  & 1.3198649  & 0.0000213  &  & 1.3199908  & 0.0000036  &  & 1.3199908  & 0.0000034 \\
   5863.64  & 1.3198674  & 0.0000195  &  & 1.3198707  & 0.0000184  &  & 1.3199915  & 0.0000030  &  & 1.3199917  & 0.0000028 \\
   5870.92  & 1.3198733  & 0.0000176  &  & 1.3198767  & 0.0000166  &  & 1.3199922  & 0.0000026  &  & 1.3199923  & 0.0000025 \\
   5878.19  & 1.3198789  & 0.0000160  &  & 1.3198822  & 0.0000151  &  & 1.3199928  & 0.0000023  &  & 1.3199930  & 0.0000022 \\
   5885.46  & 1.3198853  & 0.0000150  &  & 1.3198884  & 0.0000140  &  & 1.3199935  & 0.0000020  &  & 1.3199936  & 0.0000019 \\
   5892.73  & 1.3198919  & 0.0000155  &  & 1.3198950  & 0.0000142  &  & 1.3199942  & 0.0000017  &  & 1.3199943  & 0.0000017 \\
\hline

  	\end{tabular}}
          \end{center}
          \label{OpticalConstantsTable}
\end{table}
}







\end{document}